 \documentclass[twocolumn,nofootinbib,showkeys,superscriptaddress]{revtex4-1} 

\usepackage{amsmath}
\usepackage{graphicx}
\usepackage{dcolumn}
\usepackage{bm}
\usepackage{color}
\usepackage[normalem]{ulem}
\usepackage[colorinlistoftodos]{todonotes}
\usepackage[colorlinks=true, allcolors=blue]{hyperref}
\usepackage{epstopdf}
\usepackage[english]{babel}
\usepackage[utf8x]{inputenc}
\usepackage[T1]{fontenc}
\usepackage{tikz}

\begin{document}

\title{Cross-sectional Urban Scaling Fails in Predicting Temporal Growth of Cities}

\author{Gang Xu}
\email[First author: ]{xugang@whu.edu.cn}
\affiliation{School of Remote Sensing and Information Engineering, Wuhan University, Wuhan 430079, China}

\author{Zhengzi Zhou}
\affiliation{School of Resource and Environmental Sciences, Wuhan University, Wuhan 430079, China}

\author{Limin Jiao}
\email[Corresponding author: ]{lmjiao@whu.edu.cn}
\affiliation{School of Resource and Environmental Sciences, Wuhan University, Wuhan 430079, China}

\author{Ting Dong}
\affiliation{School of Resource and Environmental Sciences, Wuhan University, Wuhan 430079, China}

\author{Ruiqi Li}
\affiliation{College of Information Science and Technology, Beijing University of Chemical Technology, Beijing 100029, China}

\date{Version \today}

\begin{abstract}
Numerous urban indicators scale with population in a power law across cities, but whether the cross-sectional scaling law is applicable to the temporal growth of individual cities is unclear. 
Here we first find two paradoxical scaling relationships that urban built-up area sub-linearly scales with population across cities, but super-linearly scales with population over time in most individual cities because urban land expands faster than population grows. 
Different cities have diverse temporal scaling exponents and one city even has opposite temporal scaling regimes during two periods, strongly supporting the absence of single temporal scaling and further illustrating the failure of cross-sectional urban scaling in predicting temporal growth of cities. 
We propose a conceptual model that can clarify the essential difference and also connections between the cross-sectional scaling law and temporal trajectories of cities. 
Our model shows that cities have an extra growth of built-up area over time besides the supposed growth predicted by the cross-sectional scaling law. Disparities of extra growth among different-sized cities change the cross-sectional scaling exponent. 
Further analyses of GDP and other indicators confirm the contradiction between cross-sectional and temporal scaling relationships and the validity of the conceptual model. Our findings may open a new avenue towards the science of cities.

\end{abstract}


\maketitle
\section*{Introduction}

Cities, residing more than half of the world’s population and continuing to grow, are of great significance in global sustainability \cite{bettencourt_evolution_2011,christa_brelsford_heterogeneity_2017}. 
With cities rising and bulk of urban data emerging, the appeal to the \textit{new Science of Cities} is getting stronger to understand our cities in depth\cite{batty_building_2012,batty_new_2013,bettencourt_towards_2019,barthelemy_modeling_2019,barthelemy_statistical_2019}. 
Scaling law is one of the general rule behind the complex system of cities, which describes how two or more attributes are functionally related \cite{west_scale:_2017,batty_theory_2013,batty_size_2008}. 
Numerous urban indicators ($Y$) scale with urban population ($N$) in a power law form ($Y_{t}=Y_0 N_{t}^\beta$) across cities, where $Y_0$ is a constant and $\beta$ is the \textit{scaling exponent}\cite{bettencourt_growth_2007}. 
Urban infrastructure related indicators (e.g., road length, gas stations) sub-linearly ($\beta<1$) scale with urban population because of \textit{economies of scale}, while social interactions related indicators (e.g., GDP, innovation, crime) super-linearly ($\beta>1$) scale with urban population due to the \textit{increasing returns} to population size, and urban indicators related to individual needs linearly ($\beta=1$) scale with population size of cities \cite{bettencourt_growth_2007}.

Since Luis Bettencourt and his colleagues opened up the field of urban scaling \cite{bettencourt_growth_2007,luis_m._a._bettencourt_origins_2013,bettencourt_unified_2010}, scholars from different disciplines have ignited heated discussion and extensive research on urban scaling in just a decade \cite{rybski_scaling_2009,um_scaling_2009,gomez-lievano_explaining_2016,ruiqi_li_simple_2017,depersin_global_2018,keuschnigg_urban_2019}. 
These research progresses on urban scaling can be divided into four aspects at least. 
(1) \textit{Verification of urban scaling}. The urban scaling has been tested in different regions, such as Europe \cite{van_raan_urban_2016,bettencourt_urban_2016}, Brazil \cite{meirelles_evolution_2018}, India \cite{sahasranaman_urban_2019} and China \cite{zund_growth_2019}. 
In addition to modern cities, urban scaling law also exists in the historical urban system \cite{ortman_pre-history_2014,ortman_settlement_2015,cesaretti_population-area_2016}. 
These verifications across regions and beyond time strongly demonstrate the universality of urban scaling law that is not restricted by history, geography and culture. 
(2)  \textit{Mechanism of urban scaling}. Network is the backbone behind all complex systems, such as the vascular network of life systems, infrastructure and social networks of urban systems \cite{west_scale:_2017}. Previous studies abstracted and built entity or virtual networks of cities, trying to explain mechanisms of urban scaling \cite{luis_m._a._bettencourt_origins_2013,ruiqi_li_simple_2017,yakubo_superlinear_2014,zhang_scaling_2015,ribeiro_model_2017}. 
(3)  \textit{Application of urban scaling}. Urban scaling provides new ideas for urban performance evaluation rather than using per capita indicators (e.g. GDP per capita). The Scale-Adjusted Metropolitan Indicator (SAMI) defined as the deviation from urban scaling has been widely used to evaluate urban performance \cite{bettencourt_urban_2010,alves_scale-adjusted_2015,alves_distance_2013,lobo_urban_2013,sahasranaman_economic_2019}. 
(4)  \textit{Criticism on urban scaling} \cite{louf_scaling:_2014,rybski_cities_2017}. 
First, scaling exponents are unstable under different extents of cities because of the complex and fuzzy boundaries \cite{arcaute_constructing_2015,cottineau_diverse_2017,batty_defining_2011}. 
Second, the applicability of the logarithmic linear regression model is questionable \cite{gudipudi_efficient_2019,leitao_is_nodate}. 
Third, scaling exponents are affected by other external conditions such as macroeconomic constitutions and public policies \cite{strano_rich_2016,muller_does_2017}. 
It is worth emphasizing that these introspections or criticisms do not prevent the scaling law from being one of the important quantitative laws of the new science of cities.

Cities are growing and systems of cities are also evolving over time \cite{paulus_evolutionary_2006,pumain_urban_2012}. 
Previous studies mainly focused on the scaling law across cities (cross-sectional scaling), the scaling law for the temporal development of a single city (temporal scaling) is rarely discussed \cite{marshall_urban_2007}. 
People take it for granted that the temporal development of the city will also follow the cross-sectional scaling law of the urban system. However, whether the cross-sectional scaling law is applicable to individual cities remains unclear. 
In recent years, a small number of studies investigated the temporal scaling and proved differences between cross-sectional and temporal scaling relationships \cite{depersin_global_2018,keuschnigg_scaling_2019,bettencourt_interpretation_2019,ribeiro_relation_2019}.

The scientific question of this study is whether there is a scaling law for the temporal development of a single city, and how cross-sectional and temporal scaling relationships interrelate in an evolving system of cities.
We first reveal the paradoxical scaling relationships of built-up area and population across cities and over time (2000-2016) in 275 Chinese cities. We build a conceptual model to show how temporal trajectories of individual cities influence the scaling exponent across cities. 
We further use gross domestic product (GDP), household electricity consumption, and road length in Chinese cities and congestion-induced delay in American cities to verify the contradiction between cross-sectional and temporal scaling relationships, and the validity of reconciling the conflict by our conceptual model.


\section*{Results}
\subsection*{Paradoxical scaling regimes}
\begin{figure*}[!ht]
	\centering
	\includegraphics[width=0.7\textwidth]{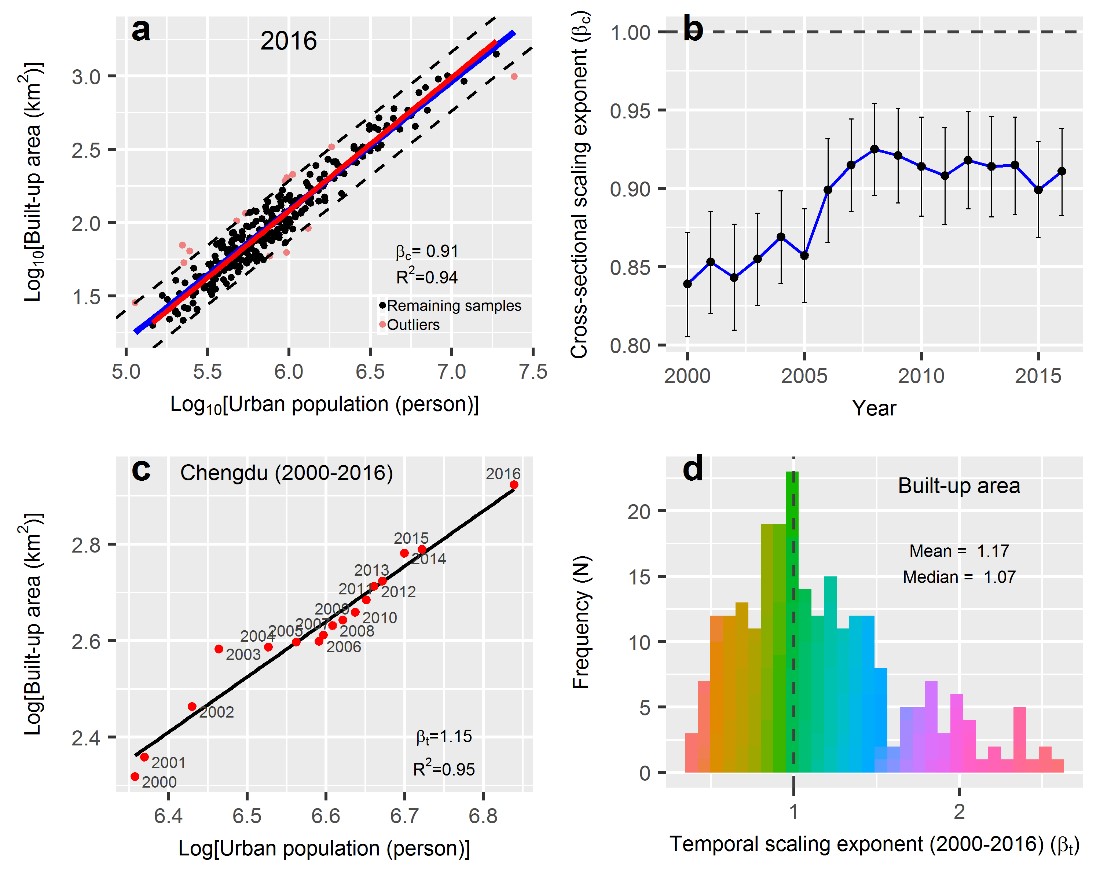}
	\caption{Paradoxical scaling regimes of built-up area and population across cities and over time. (a) Cross-sectional scaling law of built-up area and population across Chinese cities in 2016. We first use OLS to fit the original data, resulting in the regression line in blue. Cities with residuals greater than twice the standard deviation of the residual are considered as outliers (outside the two dashed lines). The OLS is used again to fit the remaining samples after removing outliers, resulting in the final cross-sectional scaling exponent ($\beta_c =0.91$). (b) The variations of $\beta_c$ with 95\% confidence intervals from 2000 to 2016. (c) Temporal scaling relationship between built-up area and population in Chengdu from 2000 to 2016 with a temporal scaling exponent $\beta_t$ of 1.15. (d) Frequency distribution of $\beta_t$ of each city from 2000 to 2016. Outliers of $\beta_t$ below $5^{th}$ percentiles and above $95^{th}$ percentiles are not shown.
		\label{fig::paradoxicalCityTime}}
\end{figure*}

We collected urban built-up area ($A$) and population ($P$) from 2000 to 2016 of 275 prefecture-level cities in China (Fig.~\ref{fig::studyArea}). We define the scaling relationship between built-up area and population across cities as the \textit{cross-sectional scaling} ($A=\alpha P^{\beta c}$) and that over time in an individual city as the \textit{temporal scaling} ($A_t=\alpha P_t^{\beta_t}$)\cite{marshall_urban_2007}. The two scaling relationships lead to the cross-sectional scaling exponent ($\beta_c$) and temporal scaling exponent ($ \beta_t$), respectively. Cross-sectionally, the built-up area sub-linearly scales with population in Chinese cities (Fig.~\ref{fig::paradoxicalCityTime}a, See scatter plots in each year from 2000 to 2016 in Fig.~\ref{fig::scaligEachYear}). The cross-sectional exponents ($\beta_c$) are all significantly less than one, varying from 0.84 to 0.93, with a mean of 0.89 (Fig.~\ref{fig::paradoxicalCityTime}b). $\beta_c$ experiences two periods before and after 2008: a persistent increase from 2000 to 2008 and a moderate decrease from 2008 to 2016.

Taking Chengdu, a megacity located in the Southwestern China, as an example, the power law ($A_t=\alpha P_t^\beta$) can also fit the relationship between built-up area and population over time \cite{marshall_urban_2007} (Fig.~\ref{fig::paradoxicalCityTime}c), and the temporal scaling exponent is greater than one ($\beta_t=1.15$). 
We fit the temporal relationships between the logarithmic forms of built-up area and population in each city from 2000 to 2016. 
The $R^2$ of the linear regression for temporal scaling varies among cities with a mean of 0.82 (Fig.~\ref{fig::frequencyR2}). 
Different cities have diverse $\beta_t$ or even opposite scaling regimes. The range of all $\beta_t$ of 275 cities is [-6.23, 13.53]. 
Values of $\beta_t$ below $5^{th}$ percentiles or above $85^{th}$ percentiles are removed, and the frequency distribution of $\beta_t$ of the remaining 248 cities is shown in Fig.~\ref{fig::paradoxicalCityTime}d. Nearly 60\% cities have a $\beta_t$ greater than 1 and the mean of $\beta_t$ of 248 cities is 1.17. Cities with a $\beta_t$ higher than 1 experience a higher rate of urban expansion than population growth over time, which has been widely evidenced\cite{marshall_urban_2007,seto_meta-analysis_2011}.

Now we come to a paradox that urban built-up area \textbf{sub-linearly} scales with population across cities at time $t$ \cite{luis_m._a._bettencourt_origins_2013,ortman_settlement_2015,batty_defining_2011}, while built-up area \textbf{super-linearly} correlated with population in more individual cities over time because urban land expands faster than population grows \cite{marshall_urban_2007,seto_meta-analysis_2011,angel_persistent_2010,xu_how_2019}. This is not just a difference in numerical values of scaling exponents \cite{depersin_global_2018,cottineau_diverse_2017,keuschnigg_scaling_2019,bettencourt_interpretation_2019,ribeiro_relation_2019}, but a difference in scaling regimes (sub-linear versus super-linear). What is the reasonable interpretation behind the paradoxical scaling regimes along the evolution of the system of cities?

\subsection*{Relationship between cross-sectional and temporal scaling}
\begin{figure*}[!ht]
	\centering
	\includegraphics[width=0.9\textwidth]{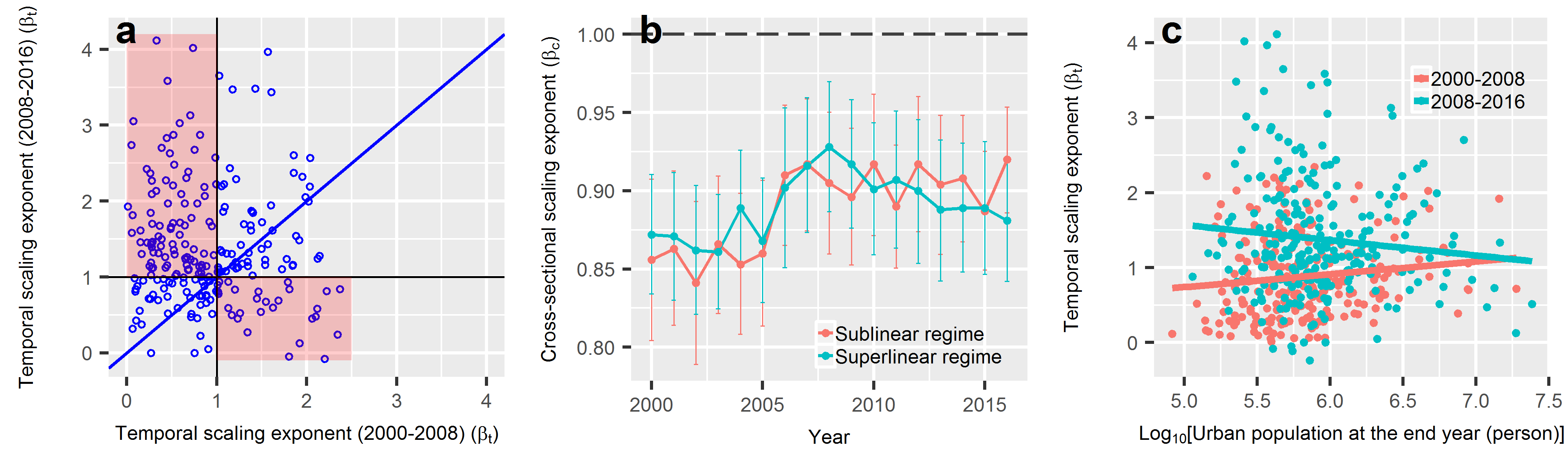}
	\caption{Relationships between cross-sectional and temporal scaling. (a) Comparisons of temporal scaling exponents ($\beta_t$) in two periods (2000-2008 and 2008-2016). Cities with a $\beta_t$ between 5th and 95th percentile are remaining, resulting in 233 cities with pairs of $\beta_t$ in the two periods. (b) The varying trend of cross-sectional scaling exponents ($\beta_c$) in two sub-systems of cities with a sub-linear or super-linear temporal scaling regime, respectively. (c) Correlations between $\beta_t$ of cities in two periods (2000-2008 and 2008-2016) and their urban population at the end year (2008 and 2016).
		\label{fig::relationCrossTemp}}
\end{figure*}

According to results in Fig.~\ref{fig::paradoxicalCityTime}b, we use the year of 2008 as the cut-off point to calculate $\beta_t$ of each city before and after this time point. There are 38\% and 64\% cities experiencing a super-linear temporal scaling regime ($\beta_t>1$) between built-up area and population during 2000-2008 and during 2008-2016, respectively (Fig.~\ref{fig::relationCrossTemp}a). The scatter plots of $\beta_t$ during the two periods are dispersed without an obvious linear trend and $\beta_t$ changes in different time periods, particularly changes through scaling regimes (cities in the red shadow in Fig.~\ref{fig::relationCrossTemp}a). This indicates that $\beta_t$ is fluctuant and not robust in an individual city.

We further test whether different temporal scaling regimes influence or change the cross-sectional scaling law. We divide cities into two sub-systems: cities with super-linear ($\beta_t>1$) and sub-linear ($\beta_t<1$) temporal scaling regimes from 2000 to 2016. Cities with $\beta_t>1$ are on the right of the dashed line in Fig.~\ref{fig::paradoxicalCityTime}d. 
Then, we calculate $\beta_c$ of the two sub-systems in each year from 2000 to 2016. Cities in the two sub-systems experience opposite temporal scaling regimes but the two sub-systems share a similar varying trend of $\beta_t$ and there are no apparent differences of $\beta_t$ between them (Fig.~\ref{fig::relationCrossTemp}b). This indicates that the absolute size of $\beta_t$ of cities does not change $\beta_c$ of the system of cities. Most importantly, $\beta_c$ of the two sub-systems is always less than 1 from 2000 to 2016, revealing the robustness of the sub-linear cross-sectional scaling law between built-up area and population.

Different cities have diverse values of $\beta_t$. Whether the $\beta_t$ of a city relies on its population size? From 2000 to 2008, $\beta_t$ of a city is positively correlated with population size at the end year of 2008, while $\beta_t$ of a city from 2008 to 2016 is negatively correlated with population size at the end year of 2016 (Fig.~\ref{fig::relationCrossTemp}c). There seems to be no consistent correlations between $\beta_t$ and population size. However, when we link their correlations to the change of $\beta_c$ during the two periods (2000-2008 and 2008-2016) (Fig.~\ref{fig::paradoxicalCityTime}d), this provides us with a thought that whether the relative size of $\beta_t$ among different-sized cities changes $\beta_c$ of the system of cities.

\subsection*{A conceptual model}
\begin{figure*}[!ht]
	\centering
	\includegraphics[width=0.5\textwidth]{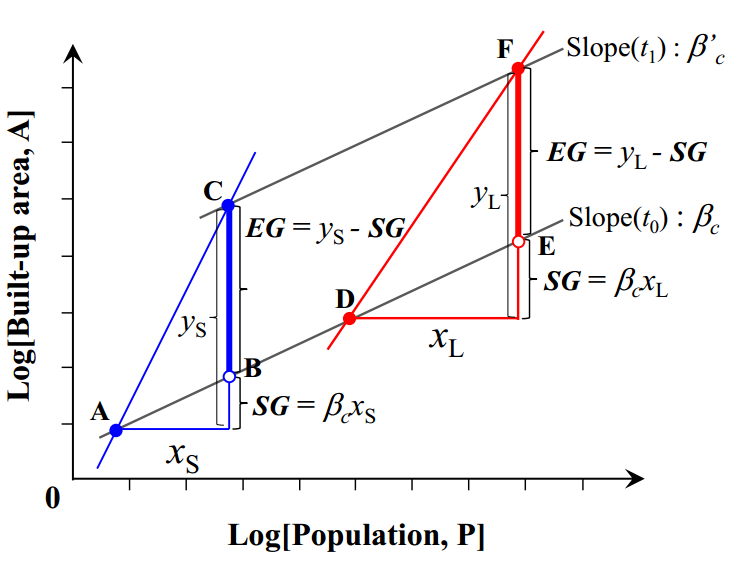}
	\caption{A conceptual model showing how temporal trajectories of individual cities influence the variation of the cross-sectional scaling exponent ($\beta_c$) from $t_0$ to $t_1$. The system of cities is generalized as two cities, a small city (in blue) and a large city (in red). The x-axis increments of urban population from $t_0$ to $t_1$ are $x_S$ and $x_L$ of the small and large cities, respectively, and their built-up areas increase $y_S$ and $y_L$, respectively. All the x-axis and y-axis increments are in the logarithm scale. The y-axis increment predicted by the cross-sectional scaling exponent of time $t_0$ is defined as the supposed growth (SG), while the difference of the actual y-axis increment ($y_S$ or $y_L$) and the supposed growth is defined as the extra growth (EG). The relative size of the extra growth among different-sized cities determines the variation of $\beta_c$ from $t_0$ to $t_1$, while it has no correlations with supposed growth.		
		\label{fig::concepDiagram}}
\end{figure*}

We build a conceptual model to present the self-consistency of cross-sectional scaling law and temporal growth of individual cities and how dynamics of individual cities ($\beta_t$) influence the variation of $\beta_c$ (Fig.~\ref{fig::concepDiagram}). We simplify the system of cities into two cities, the small city (in blue) and the large city (in red). The cross-sectional scaling exponents at time $t_0$ and $t_1$ are $\beta_c$ (<1) and$\beta_c^{,}$, respectively. From $t_0$ to $t_1$, their population sizes in the logarithm scale increase $x_S$ and $x_L$ of the small and large cities, respectively, and their built-up areas increase $y_S$ and $y_L$, respectively. Taking $x_S$ as an example, its calculation is as follows:
\begin{equation}
x_S = \log P_1 - \log P_0= \log \frac{P1}{P0} \quad.
\label{eq::3}
\end{equation} 
where $P_0$ and $P_1$ are the urban population of the small city at $t_0$ and $t_1$, respectively. We can calculate $y_S$, $x_L$, and $y_L$ in the same way.

If the small city grows over time following the scaling law across cities ($\beta_c$), then it moves from A to B (Fig.~\ref{fig::concepDiagram}). Under this circumstance, the y-axis increment equals to $\beta_c x_S$, which is defined as the \textit{supposed growth} (SG) predicted by the cross-sectional scaling law. However, the temporal trajectory of an individual city ($\beta_t$) usually violates the cross-sectional scaling law ($\beta_c$). The final position of the small city is not B, but C. We define the increment from B to C as the \textit{extra growth} (EG), which can be calculated as follows:
\begin{equation}
EG=y_S-SG=y_s-\beta_c x_S \quad.
\label{eq::4}
\end{equation} 

The introduction of the extra growth explains the paradoxical regimes and unifies the scaling law across cities and temporal trajectories of individual cities. If the large city has a higher extra growth than the small city, $\beta_t^{,}$ increases at time $t_1$; otherwise,$\beta_c^{,}$ decreases. From the two cites back to the system of cities, we can interpret that it is the differences of the  extra growth among different-sized cities that determines the change of  $\beta_c$ from $t_0$ to $t_1$, while the supposed growth does not influence. Specifically, if the extra growth of built-up area from $t_0$ to $t_1$ of a city is positively correlated with its population size at $t_1$, the cross-sectional scaling exponent ($\beta_c$) increases from $t_0$ to $t_1$; otherwise,$\beta_c$ decreases.

\begin{figure*}[!ht]
	\centering
	\includegraphics[width=0.75\textwidth]{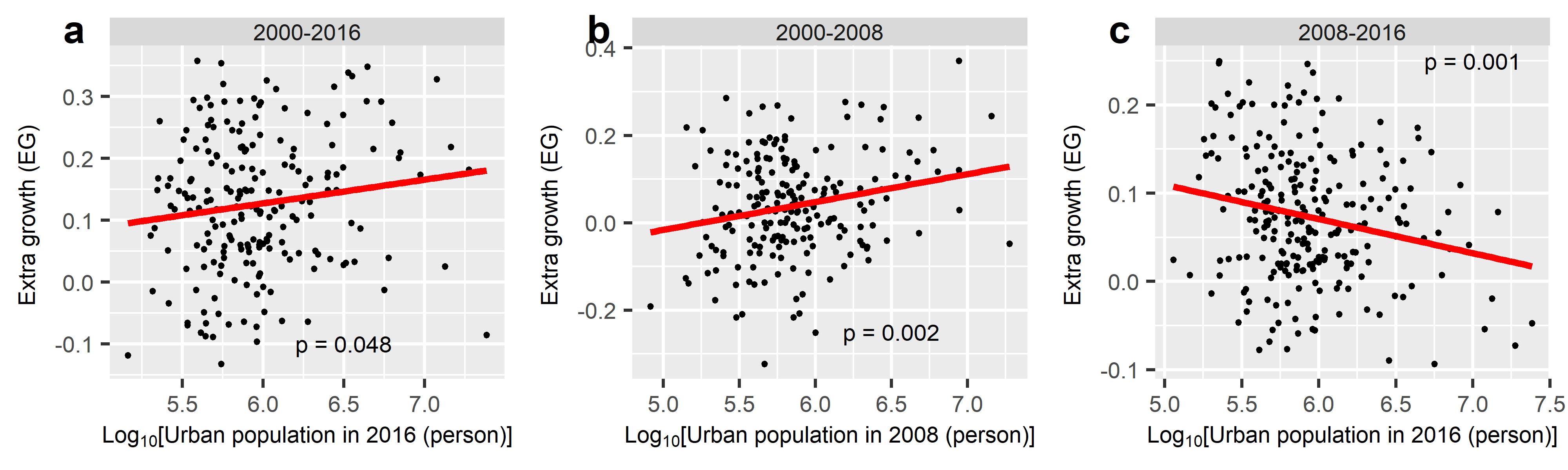}
	\caption{Scatter plots of extra growth of built-up area from $t_0$ to $t_1$ and urban population at time $t_1$. (a) From 2000 to 2016, (b) From 2000 to 2008, and (c) From 2008 to 2016. The extra growth and urban population are both in the logarithm scale.	
		\label{fig::scatterPlotExcess}}
\end{figure*}

Our empirical analyses confirm with the interpretation of the conceptual model(Fig.~\ref{fig::scatterPlotExcess}). The extra growth of built-up area from 2000 to 2016 is significantly and positively correlated with the urban population in 2016 (Fig.~\ref{fig::scatterPlotExcess}a), which indicates that large cities have higher extra growth, finally resulting in the increase of $\beta_c$ during this period (Fig.~\ref{fig::paradoxicalCityTime}b). Specific to the two sub-periods, the positive correlation between them corresponds the increase of $\beta_c$ during 2000-2008 (Fig.~\ref{fig::scatterPlotExcess}b); while the negative correlation explains the decrease of $\beta_c$ during 2008-2016 (Fig.~\ref{fig::scatterPlotExcess}c). Their correlations are all significant at the level of 0.05 (p < 0.05).

\subsection*{Verification using other indicators}
\begin{figure*}[!ht]
	\centering
	\includegraphics[width=0.8\textwidth]{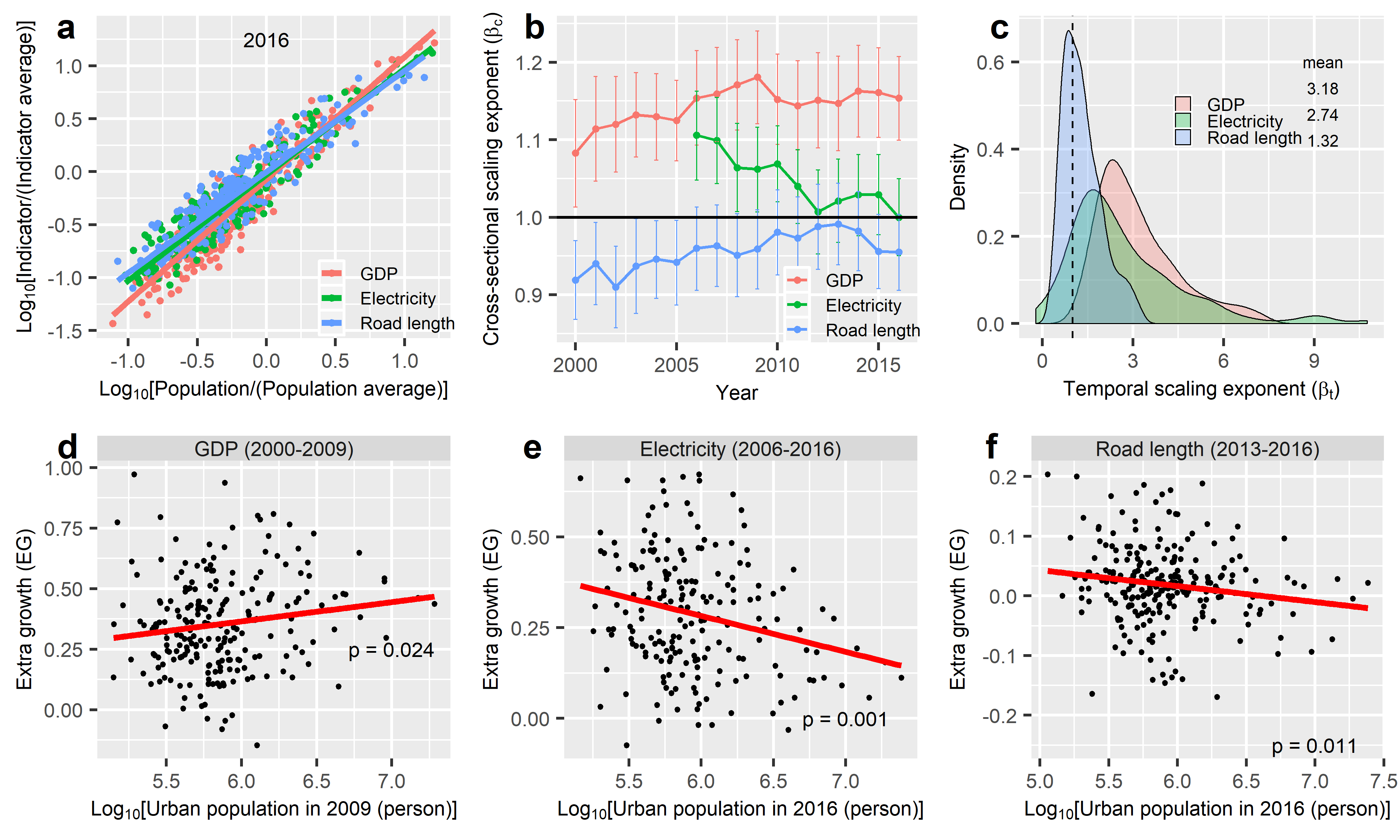}
	\caption{Verification of the paradoxical scaling regimes across cities and over time and the validity of the conceptual model using GDP, household electricity consumption, and road length in 275 Chinese cities. (a) Scaling relationships between three urban indicators and population in 2016. Urban indicators and urban population are divided by their averages, respectively. (b) Varying trends of the cross-sectional scaling exponents ($\beta_c$). (c) The distribution of probability density of temporal scaling exponents ($\beta_t$). (d) Scatter plots of extra growth of GDP from 2000 to 2009 and population in 2009. (e) Scatter plots of extra growth of household electricity consumption from 2006 to 2016 and population in 2016. (f) Scatter plots of extra growth of road length from 2013 to 2016 and population in 2016.	
		\label{fig::verificationRegimes}}
\end{figure*}

We first use three other urban indicators (GDP, household electricity consumption, and road length) in Chinese cities to verify our findings. The power law holds well between urban indicators and population across cities in each year from 2000 to 2016 (Fig.~\ref{fig::verificationRegimes}a, Fig.~\ref{fig::otherIndicators}, the time period of household electricity consumption is from 2006 to 2016). The $\beta_c$ of GDP is all significantly greater than 1, indicating a super-linear cross-sectional scaling regime all the time (Fig.~\ref{fig::verificationRegimes}b). Household electricity consumption, strongly related to individual needs, is expected to linearly scale with population size, but its $\beta_c$ is significantly greater than 1 from 2006 to 2010. Urban road, as a kind of infrastructure, its length is expected to sub-linearly scale with population size, but the $\beta_c$ of road length is not significantly less than 1 in most years.
We calculate $\beta_t$ for GDP and road length of each city from 2000 to 2016 and $\beta_t$ for electricity consumption from 2006 to 2016. The curves of probability density (which is a smoothed version of the histogram) of $\beta_t$ are presented in Fig.~\ref{fig::verificationRegimes}c. No matter of what kind of urban indicators, most cities have a $\beta_t$ greater than 1 (the vertical dashed line), even though the $\beta_t$ for road length, whose $\beta_c$ is less than 1. The average $\beta_t$ of cities for GDP, electricity consumption, and road length are 3.18, 2.74, and 1.32, respectively. Correspondingly, the average $\beta_c$ for GDP, electricity consumption, and road length are 1.14, 1.05, and 0.96, respectively. The $\beta_c$ and $\beta_t$ are completely different from each other, strongly confirming the finding of paradoxical scaling regimes across cities and over time.

The relationship between the extra growth and population size explains the change of $\beta_c$ of GDP, electricity consumption, and road length (Fig.~\ref{fig::verificationRegimes}b-f). From 2000 to 2009, the extra growth of GDP is significantly and positively correlated with population size in 2009, which means large cities have higher extra growth of GDP, resulting in the increase of $\beta_c$ of GDP during 2000-2009 (Fig.~\ref{fig::verificationRegimes}d). The $\beta_c$ of electricity obviously decreases from 2006 to 2016, which can be explained by the significant and negative correlation between the extra growth and population size (Fig.~\ref{fig::verificationRegimes}e). The extra growth of road length from 2013 to 2016 are significantly and negatively correlated with population size in 2016, which is consistent with the decrease of $\beta_c$ of road length during 2013-2016 (Fig.~\ref{fig::verificationRegimes}f). These correlations confirm our second finding that the extra growth of different-sized cities determines the change of $\beta_c$.

\begin{figure*}[!ht]
	\centering
	\includegraphics[width=0.8\textwidth]{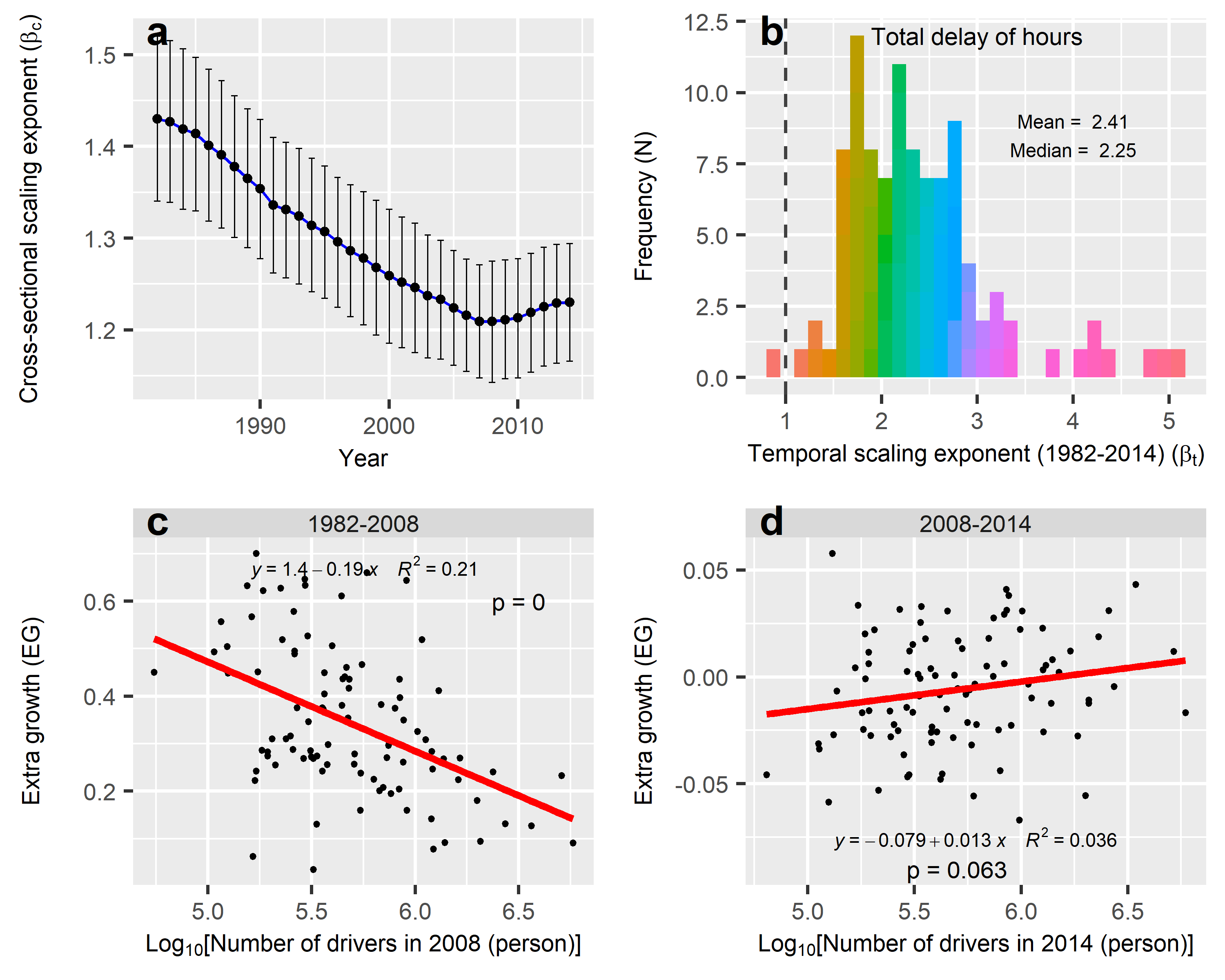}
	\caption{Verification of the paradoxical scaling regimes across cities and over time and the validity of the conceptual model using congestion-induced delay in 101 American cities. (a) Varying trends of the cross-sectional scaling exponents ($\beta_c$) between congestion-induced delay and the number of car commuters from 1982 to 2014. (b) Frequency distribution of $\beta_t$ from 1982 to 2014 of each city (101 cities). (c) Scatter plots of extra growth of congestion-induced delay from 1982 to 2008 and population in 2008 in both logarithm scales. (d) Scatter plots of extra growth of congestion-induced delay from 2008 to 2014 and the number of car commuters in 2014 in both logarithm scales.	
		\label{fig::verificationRegimesAmer}}
\end{figure*}
We further use the congestion-induced delay in 101 American cities from 1982 to 2014 to verify our findings (Fig.~\ref{fig::verificationRegimesAmer}). This is a public-accessed dataset and it has been used in urban scaling studies \cite{depersin_global_2018,bettencourt_interpretation_2019}. The $\beta_c$ of congestion-induced delay (hours) and the number of car commuters (representing population size) decreases from 1982 to 2008, and then it increases till 2014 (Fig.~\ref{fig::verificationRegimesAmer}a). The $\beta_c$ varies from 1.21 to 1.43, with a mean of 1.29. Different cities have diverse $\beta_t$ during 1982-2014, with a mean of 2.41, which are definitely different from $\beta_c$ (Fig.~\ref{fig::verificationRegimesAmer}b). The correlation between extra growth and population size explains the decreases of $\beta_c$ from 1982 to 2008 and its increase from 2008 to 2014, verifying the validity of the conceptual model (Fig.~\ref{fig::verificationRegimesAmer}c-d).

\section*{Discussion}
We discovered the paradoxical scaling regimes of built-up area and population across cities and over time. Built-up area sub-linearly scales with population across cities at time t \cite{luis_m._a._bettencourt_origins_2013,rybski_cities_2017,batty_defining_2011}. Temporally, the worldwide faster growth rate of built-up area than population in most individual cities results in a super-linear scaling relationship over time \cite{marshall_urban_2007,seto_meta-analysis_2011,angel_persistent_2010,xu_how_2019}. The paradoxical scaling regimes imply that scaling exponent found across cities cannot be applied to individual cities\cite{depersin_global_2018}. Previous studies also have found different scaling exponents or regimes caused by varying definitions of cities\cite{arcaute_constructing_2015,cottineau_diverse_2017,louf_scaling:_2014} or different groups \cite{strano_rich_2016}. Different from these studies, the paradoxical scaling regimes found in this study reflects the difference of characteristics across cities and over time along the evolving system of cities.

Our study illustrates the absence of single scaling law for the temporal development of a single city from the following three considerations. 
(1) In terms of certain urban indicator, different cities have diverse temporal scaling exponents and even opposite scaling regimes (Fig.~\ref{fig::paradoxicalCityTime}d, Fig. ~\ref{fig::verificationRegimes}c, Fig. ~\ref{fig::verificationRegimesAmer}b ). 
(2) One city has varying scaling exponents during different periods (Fig.~\ref{fig::relationCrossTemp}a). 
(3) There is no obvious linear relationship between urban indicators and population size over time in some cities, resulting in the bad performance of the regression model (Fig.~\ref{fig::frequencyR2}). 
These facts strongly and powerfully demonstrate the temporal growth of individual cities will not follow the scaling law across cities and the cross-sectional scaling law cannot be applied to individual cities. Compared with the stable urban scaling law of urban systems, there is no uniform and consistent law for the temporal trajectory of a single city, particularly in countries with rapid urban transformation.

\vspace{10pt}
We proposed a conceptual model to reveal the evolutionary relationship between $\beta_c$ and $\beta_t$. We found that it is the relative size, not the absolute size, of $\beta_t$ of different-sized cities influences the variation of $\beta_c$ of the system of cities. A city has an extra growth of built-up area when it grows in population, beside the supposed growth predicted by the cross-sectional scaling law. If large cities have higher extra growth, $\beta_c$ increases over time; otherwise, it decreases. The conceptual model not only reveals the contradiction between $\beta_c$ and $\beta_t$, but also unify $\beta_c$ and $\beta_t$, and further explains how temporal trajectories of individual cities ($\beta_t$) influences the scaling exponent across cities ($\beta_c$).
\vspace{10pt}

$\beta_c$ reflects characteristics of the evolving system of cities. The sub-linear scaling regime ($\beta_c$ < 1) between built-up area and population indicates that large cities have a higher urban land use efficiency compared to small cities. The comparative advantage of land use efficiency of large cities decreased from 2000 to 2008 and then generally increased in 2008-2016. The $\beta_c$ of GDP increased from 2000 to 2009 and then decreased to 2016, which indicates that large cities have a higher and higher efficiency of economy performance from 2000 to 2009, but the pioneering advantages of large cities decreased from 2009 to 2016. The variation trend of $\beta_c$ for GDP is nearly consistent with built-up area. China joined the WTO in 2001. Rapid urban expansion and fast economy growth mainly took place in large cities during the first decade of the 21st century, resulting in the decline in land use efficiency but the increase in economic efficiency in large cities. After the economy crisis in 2008, China’s economic center shifted to the interior and the inland, supporting the growth of small and medium-sized cities, leading to the increase in land use efficiency but the decline in economic efficiency in large cities. In addition, the decreasing of $\beta_c$ for electricity reflects that the electricity supply meets the demand of citizens in cities of all sizes. The $\beta_c$ of road length has an overall increasing trend from 2000 to 2013 but with a decline from 2013 to 2016, indicating that large cities have a higher growth rate of roads before 2013.

\section*{Conclusions}
Scaling law is the simple rule behind the complex system of cities. Urban area, as a kind of infrastructure, sub-linearly scales with urban population because of economies of scale. However, urban area expands faster than population grows in more individual cities over time, resulting in a super-linear relationship over time. In this study, we revealed the paradoxical scaling relationship between urban area and population across cities and over time for the first time. The paradox clearly implies that scaling laws across cities cannot be applied to the temporal growth of individual cities, which is usually taken for granted.
\vspace{15pt}

We build a conceptual model to unify the difference between the scaling relationship across cities and that over time, which shows cities have an extra growth of urban area besides the supposed growth predicted by the scaling law across cities. The differences of extra growth of urban area among different-sized cities determine the change of the scaling exponent across cities. If larger cities have higher extra growth, then the scaling exponent across cities will increase at the next time point. We further verify the paradoxical scaling relationships across cities and over time and the effectiveness of the proposed conceptual model using other urban indicators in Chinese cities and American cities. Our analyses on the scaling law across cities and over time may open a new avenue towards the science of cities.

\section*{Materials and Methods}
\subsection*{Study area and data}

Our sample cities are 275 Chinese prefecture-level cities, including megacities like Beijing and Shanghai and also medium-sized and small cities (Fig.~\ref{fig::studyArea}). We collected urban population and four urban indicators (built-up area, GDP, household electricity consumption, and road length) in those Chinese cities from the China Urban Construction Statistical Yearbook and the China City Statistical Yearbook. The time period of the data is from 2000 to 2016 in each year except for electricity consumption data, which is from 2006 to 2016.
\vspace{15pt}

A key difficulty in urban studies is finding a practical way to define the urban extent\cite{jiao_urban_2015}, which is also a skepticism and criticism point in urban scaling analysis\cite{arcaute_constructing_2015,cottineau_diverse_2017}. In this study, our statistical data of the urban extent is the built-up area, which is defined as the area where urban infrastructure (e.g. road, water supply, electricity, etc.) and public facilities (e.g. education, medicine care, public administration, etc.) are basically available. The built-up area reflects the physical extent of urbanized areas in Chinese cities. The population used in a city is urban resident population within the urbanized area, including registered population and floating population\cite{xu_how_2019}. The statistical extent (or unit) for GDP, household electricity consumption, and road length is also the urbanized area. The validation dataset of congestion-induced delay in 101 American cities from 1982 to 2014 is accessed from the Texas A\&M Transportation Institute (TTI) in the Urban Mobility Report (UMR).
\vspace{15pt}

All data sets are publicly available. Data from the \textit{China Urban Construction Statistical Yearbook} \href{http://www.mohurd.gov.cn/xytj/tjzljsxytjgb/jstjnj/index.html}{(http://www.mohurd.gov.cn/xytj/tjzljsxytjgb/jstjnj)} 
is open access on the website of the Ministry of Housing and Urban-Rural Development of China. Data from the \textit{China City Statistical Yearbook} is accessed from the website of China National Knowledge Infrastructure (CNKI) \href{http://data.cnki.net/yearbook/Single/N2018050234}{(http://data.cnki.net/yearbook/Single/N2018050234)}. Congestion-induced delay data in American cities can be accessed from the Texas A\&M Transportation Institute (TTI) \href{http://tti.tamu.edu/documents/ums/congestion-data/complete-data.xlsx}{(http://tti.tamu.edu/documents/ums/congestion-data/complete-data.xlsx)}. All data sets are available from the authors upon reasonable request.


\subsection*{Power function fitting}
Urban indicators ($Y$) scale with population size($N$) in a power law form across cities at time t:
\begin{equation}
Y_t=Y_0 N_t^\beta \quad.
\label{eq::1}
\end{equation}
where $\beta$ is the scaling exponent and $Y_0$ is the constant.
Taking the logarithm of both sides of Eqs. (\ref{eq::1}) results in a linear relationship between $\log N_t$ and $\log Y_t$.
\begin{equation}
\log Y_t=\beta \log N_t + \log Y_0 \quad.
\label{eq::2}
\end{equation} 

We use the ordinary least-squares (OLS) linear regression to fit the logarithms and the slope is the scaling exponent ($\beta$). Although this usual approach has been criticized\cite{gudipudi_efficient_2019,leitao_is_nodate}, it is still a popular approach in urban scaling analysis because of its simplicity and ease of implementation.
In spite of fitting the power law relationship between urban indicators and population size across cities at time $t$, we also adopt the linear regression model to fit the logarithms of urban indicators and population in each city over time. We define the scaling relationship between urban indicators and population across cities as the cross-sectional urban scaling and that over time in an individual city as the temporal urban scaling, resulting in the cross-sectional scaling exponent($\beta_c$)  and temporal scaling exponent($\beta_t$), respectively.

\section*{Funding}
This research was funded by the China Postdoctoral Science Foundation (BX20190251), the National Natural Science Foundation of China (41971368, 61903020).

\bibliographystyle{unsrt}

\pagebreak

\clearpage

\onecolumngrid

\section{Supplementary Information}

\begin{appendix}

\begin{figure*}[!ht]
	\centering
	\includegraphics[width=0.8\textwidth]{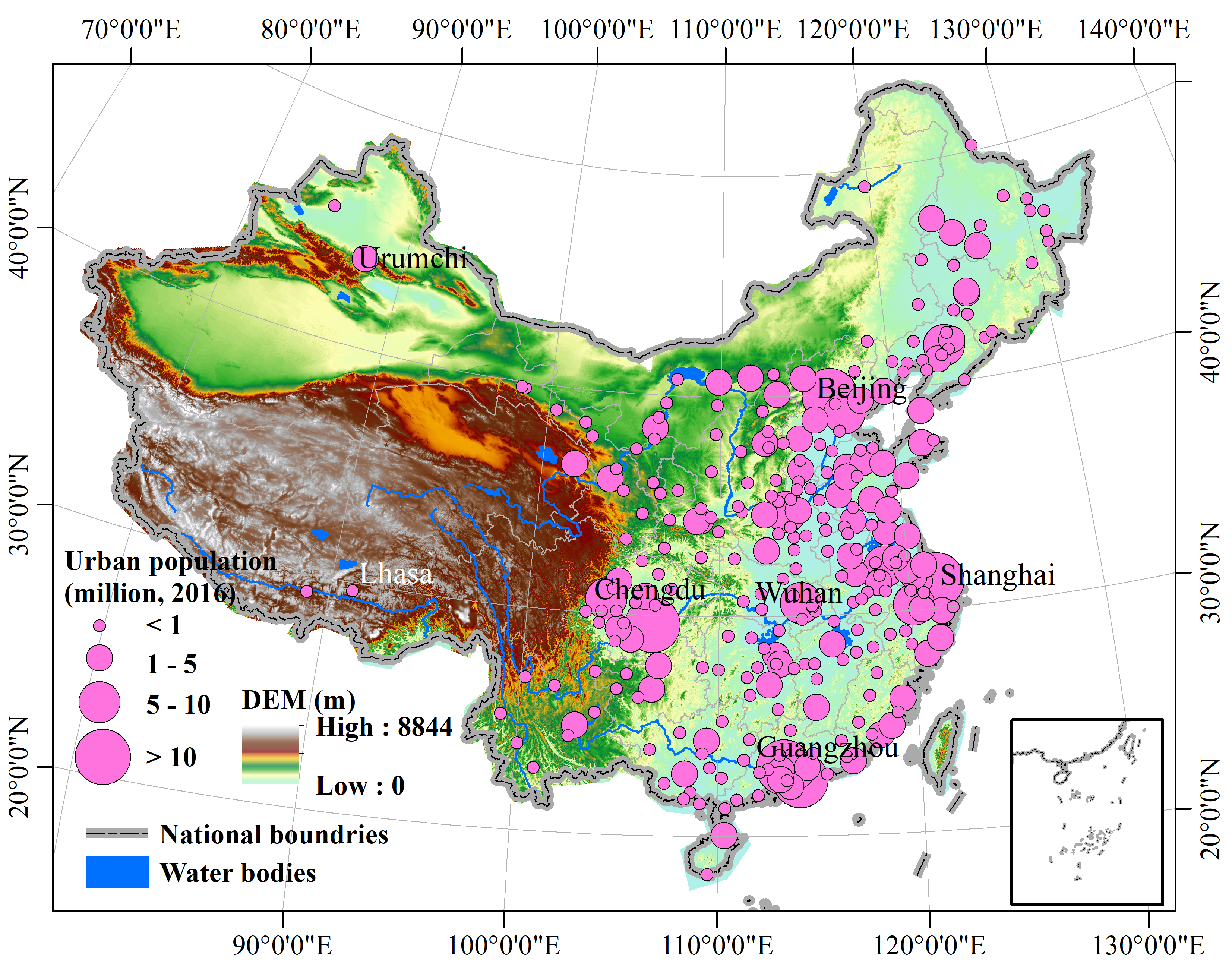}
	\caption{The spatial distribution of 275 Chinese prefecture-level cities and their population in 2016.
		\label{fig::studyArea}}
\end{figure*}

\begin{figure*}[!ht]
	\centering
	\includegraphics[width=0.9\textwidth]{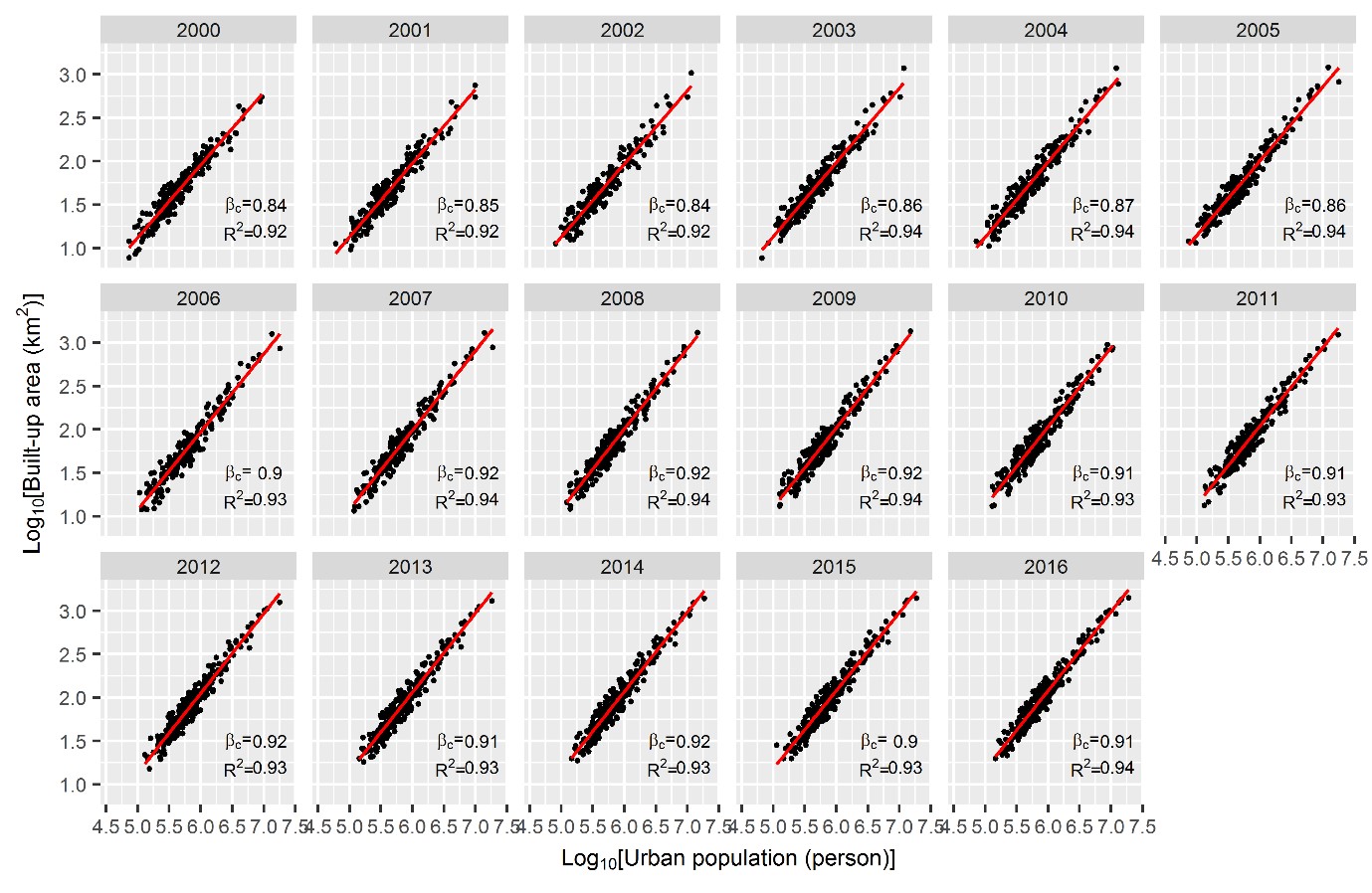}
	\caption{Scaling of built-up areas and population across cities in China in each year from 2000 to 2016. Built-up areas and population are in the logarithmic scale with a linear fit using the ordinary least squares (OLS). Cities with residuals greater than twice the standard deviation of residual in the initial regression model are considered as outliers. After the outliers are removed, the linear regression is performed again to obtain the final regression equation and scaling exponent, which are shown as the red straight lines. For all cases, p-value < 0.001.
		\label{fig::scaligEachYear}}
\end{figure*}

\begin{figure*}[!ht]
	\centering
	\includegraphics[width=0.6\textwidth]{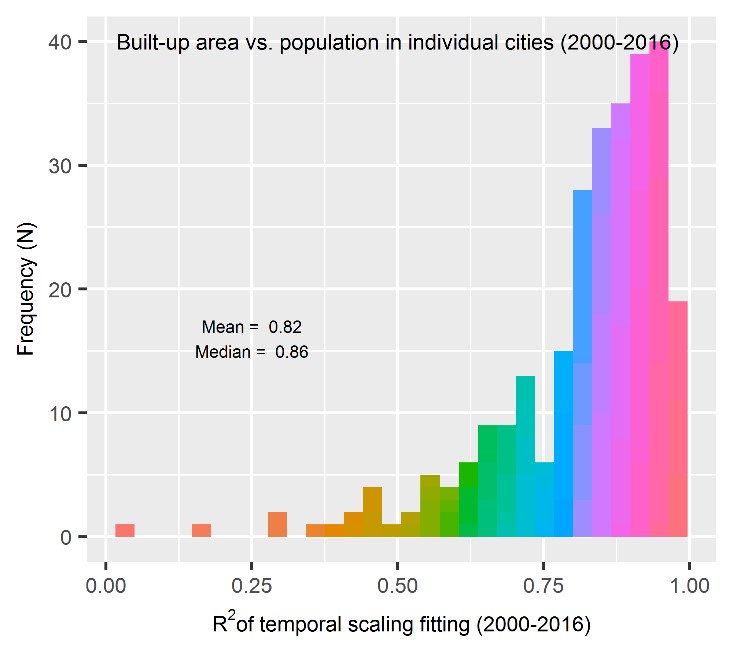}
	\caption{Frequency distribution of $R^2$ of the linear regression for fitting the temporal relationship between built-up area and urban population from 2000 to 2016 in 275 individual cities.
		\label{fig::frequencyR2}}
\end{figure*}

\begin{figure*}[!ht]
	\centering
	\includegraphics[width=0.9\textwidth]{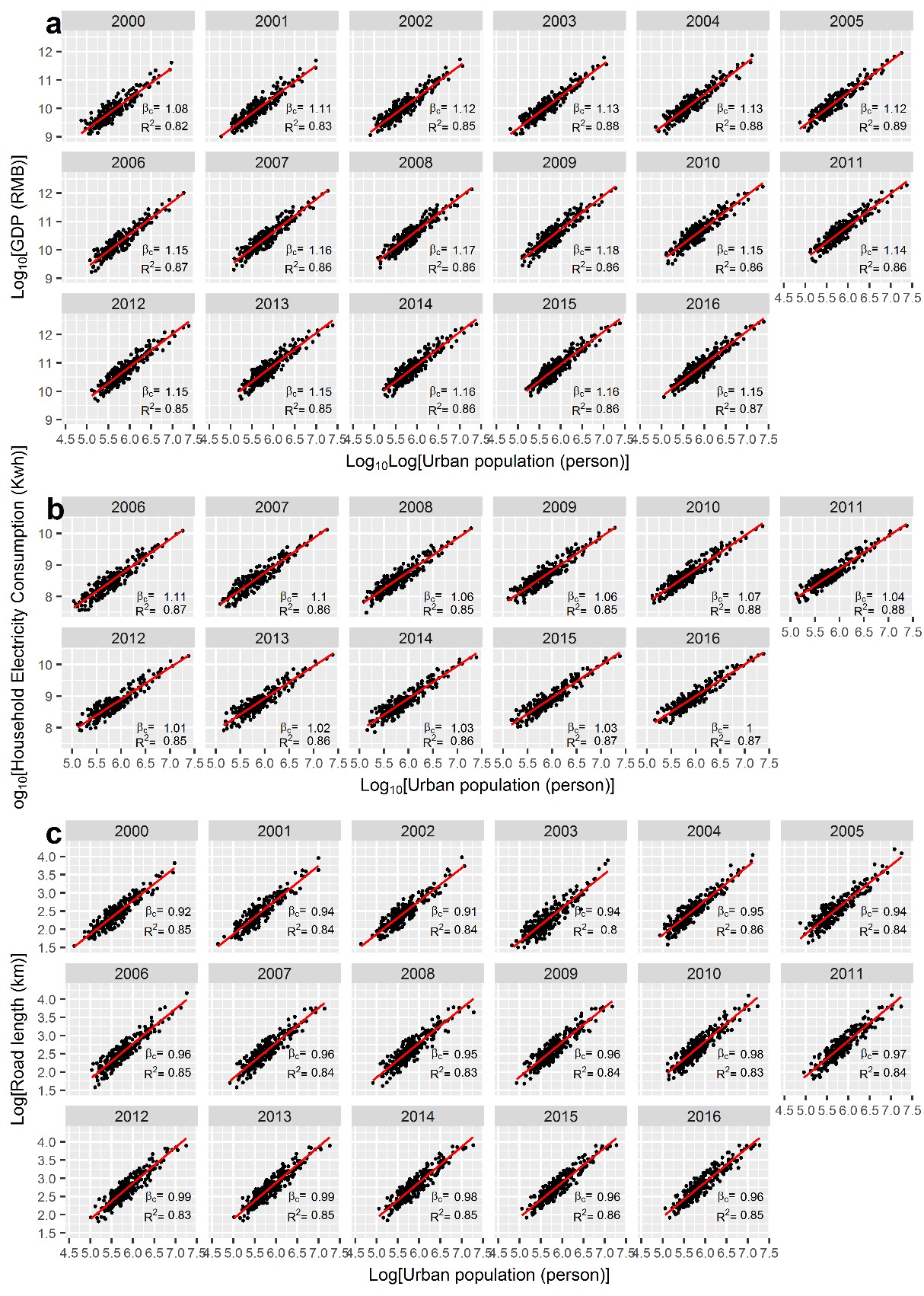}
	\caption{Scaling of three urban indicators and population across Chinese cities in each year. (a) GDP (gross domestic product), (b) Household electricity consumption, and (c) Road length. Both urban indicators and population are in the logarithmic scale with a linear fit using the ordinary least squares (OLS). Cities with residuals greater than twice the standard deviation of residual in the initial regression model are considered as outliers. After the outliers are removed, the linear regression is performed again to obtain the final regression equation and scaling exponent, which are shown as the red straight lines. For all cases, p-value < 0.001.
		\label{fig::otherIndicators}}
\end{figure*}

\end{appendix}
\end{document}